\documentclass[10pt,twocolumn,letterpaper]{article}

\usepackage{cvpr}
\usepackage{times}
\usepackage{epsfig}
\usepackage{graphicx}
\usepackage{amsmath}
\usepackage{amssymb}

\usepackage{tikz} 
\usepackage{pgfplots}
\pgfplotsset{compat=1.5}

\usepackage{subfigure}
\usepackage{caption}
\usepackage{subfigure}
\usepackage{color}
\usepackage{placeins}
\usepackage{float}
\usepackage{tabularx,colortbl}
\usepackage{times}
\usepackage{epstopdf}
\usepackage{mathrsfs}
\usepackage{diagbox}
\usepackage{multirow}
\usepackage{algorithm}
\usepackage{algorithmic}
\usepackage{galois}
\usepackage{url}

\usepackage{relsize}

\usepackage[pagebackref=true,breaklinks=true,letterpaper=true,colorlinks,bookmarks=false]{hyperref}

\cvprfinalcopy 

\def\cvprPaperID{0011} 

\ifcvprfinal\fi

\begin{document}

\title{Low Bitrate Image Compression with Discretized Gaussian Mixture Likelihoods}

\author{
Zhengxue Cheng$^{1}$, Heming Sun$^{2,3}$, Jiro Katto$^{1}$ \\
\small{$^{1}$ Department of Computer Science and Communication Engineering, Waseda University, Tokyo, Japan}\\
\small{$^{2}$ Waseda Research Institute for Science and Engineering, Tokyo, Japan $^{3}$ JST, PRESTO, 4-1-8 Honcho, Kawaguchi, Saitama, Japan}\\
{\tt\small zxcheng@asagi.waseda.jp}
}


\maketitle

\begin{abstract}
    In this paper, we provide a detailed description on our submitted method \emph{Kattolab} to Workshop and Challenge on Learned Image Compression (CLIC) 2020. Our method mainly incorporates discretized Gaussian Mixture Likelihoods to previous state-of-the-art learned compression algorithms. Besides, we also describes the acceleration strategies and bit optimization with the low-rate constraint. Experimental results have demonstrated that our approach \emph{Kattolab} achieves 0.9761 and 0.9802 in terms of MS-SSIM at the rate constraint of 0.15 \emph{bpp} during the validation phase and test phase, respectively.
\end{abstract}

\section{Introduction}

Image compression is a fundamental research topic in the field of image signal processing for many decades to achieve efficient image transmission and storage. Traditional image compression standards have been developed for a long time, such as JPEG~\cite{IEEEexample:JPEG}, JPEG2000~\cite{IEEEexample:JPEG2000}, HEVC/H.265~\cite{IEEEexample:HEVC} and ongoing Versatile Video Coding (VVC)~\cite{IEEEexample:VVC}. Typically they rely on hand-crafted creativity to present a fixed encoder/decoder (codec) block diagrams. They use predefined transform matrix, intra prediction, quantization, arithmetic coders and various post filters to reduce spatial redundancy and improve the coding efficiency. The standardization of a traditional codec has historically spanned many years. Along with the fast development of new image formats and the proliferation of high-resolution mobile devices, existing image compression standards are not expected to be an optimal and general solution for all kinds of image contents.

Recently, various approaches has been investigated for end-to-end learned image compression such as early-stage differentiable quantization for end-to-end training~\cite{IEEEexample:Theis, IEEEexample:Balle, IEEEexample:softQuan}, recurrent neural networks-based methods~\cite{IEEEexample:Toderici01, IEEEexample:Toderici, IEEEexample:Nick}, some generative models~\cite{IEEEexample:waveone, IEEEexample:MITgan, IEEEexample:Extreme}, content-weighted strategy~\cite{IEEEexample:HKPU}, conditional probability models~\cite{IEEEexample:conditional}, de-correlating different channels using principle component analysis~\cite{IEEEexample:PCS2018, IEEEexample:CLIC2018cheng}, or energy compaction based approach~\cite{IEEEexample:TMM2019cheng, IEEEexample:CVPR2019}. The most representative approaches are adaptive entropy models for rate estimation, including a hyperprior~\cite{IEEEexample:Balle2} and its variants~\cite{IEEEexample:David, IEEEexample:Lee, IEEEexample:CVPR2020} to achieve state-of-the-art performance. Specifically, the work~\cite{IEEEexample:Balle2} proposed a scale hyperprior, by encoding additional bits to build the entropy model for latent codes. The work~\cite{IEEEexample:David} jointly combined an autoregressive mask convolution and a mean-scale hyperprior to make entropy model more accurate. The work~\cite{IEEEexample:Lee} proposed a quite similar idea by considering two types of contexts, bit-consuming contexts (i.e., hyperprior) and bit-free contexts (i.e., mask convolution model) to realize a context-adaptive entropy model. The work~\cite{IEEEexample:CVPR2020} further extended the single Gaussian model to Gaussian mixture likelihoods to further improve the accuracy of entropy models. Our method is based on these recent techniques and apply them to low bitrate image compression.

In this paper, we present a detailed description on our submitted method to Workshop and Challenge on Learned Image Compression (CLIC) 2020. The network architecture combines recent techniques, including deep residual blocks, subpixel convolution and attention modules. The entropy model utilizes discretized Gaussian mixture likelihoods to achieve more accurate entropy model than single Gaussian model. Besides, we also apply some acceleration strategies and bit optimization to meet the limit of 10 hours decoding time and 0.15 \emph{bpp} rate constraint in the CLIC low-rate track. Experimental results have demonstrated that our approach \emph{Kattolab} achieves 0.9761 in terms of MS-SSIM at the rate constraint of 0.15 \emph{bpp} during the validation phase.


\section{Learned Low Bitrate Image Compression}

\subsection{Network Architecture}

\begin{figure*}[tb]
	\centerline{\psfig{figure=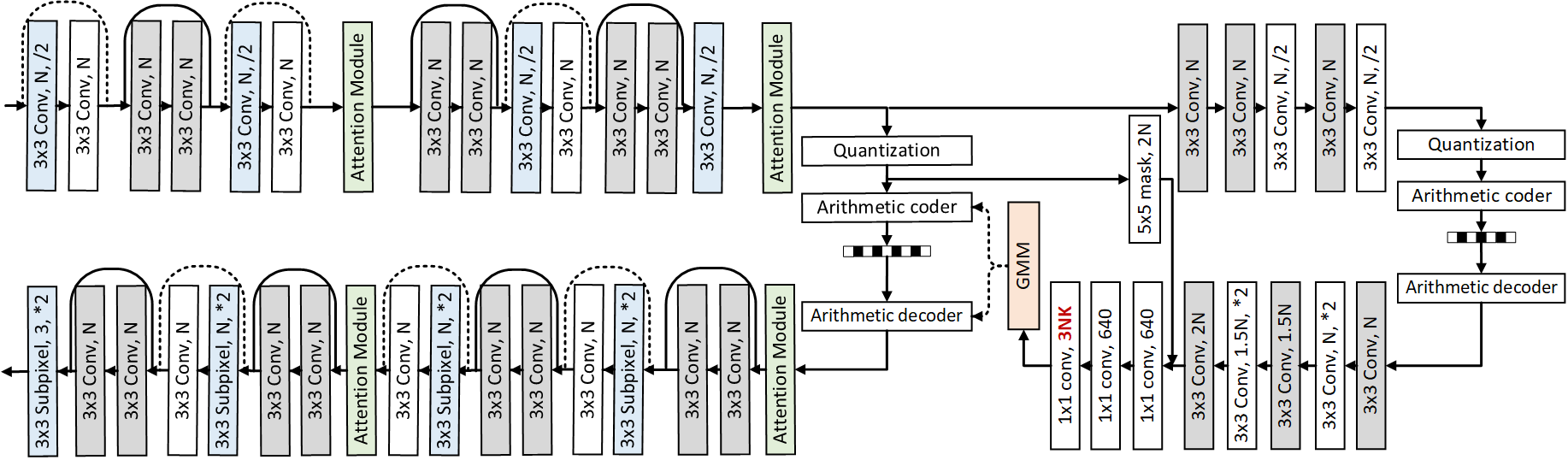,width=173mm} }
	\caption{The overall network architecture we used in \emph{Kattolab}, where the green blocks denote attention modules, the orange block denotes Gaussian mixture model ($K$ denotes the number of mixtures, $N$ denotes the number of filters) and the blue blocks represent the downsampling and upsampling units, implemented by stride-2 convolutions and sub-pixel convolutions.}
	\label{fig:net}
\end{figure*}

The network architecture we used is shown in Fig.~\ref{fig:net}, referring to~\cite{IEEEexample:CVPR2020}. Compared to the work~\cite{IEEEexample:Balle2}, the backbone network architecture has been improved by using residual blocks, subpixel convolution and attention modules. Based on the observations of~\cite{IEEEexample:CLIC2019}, deep residual blocks can achieve more larger and effective receptive field than $5\times5$ filters, therefore, we used the residual block implemented by stacked $3\times3$ filters to replace $5\times5$ filters as downsampling units at the encoder side and mirrored them at the decoder. GDN and IGDN~\cite{IEEEexample:Balle3} are only followed by the convolution with the stride of 2, and ReLU is used after other convolution filters. Besides, \cite{IEEEexample:CLIC2019} also found subpixel convolution could maintain more details compared to transposed convolution to improve the quality of reconstructed images, so we used subpixel convolution to upsample the feature sizes at the decoder side.

\begin{figure}[tb]
\subfigure[The attention module, used in~\cite{IEEEexample:HaojieLiu}]{
\label{Fig.attention.1}
\includegraphics[width = 6.5cm]{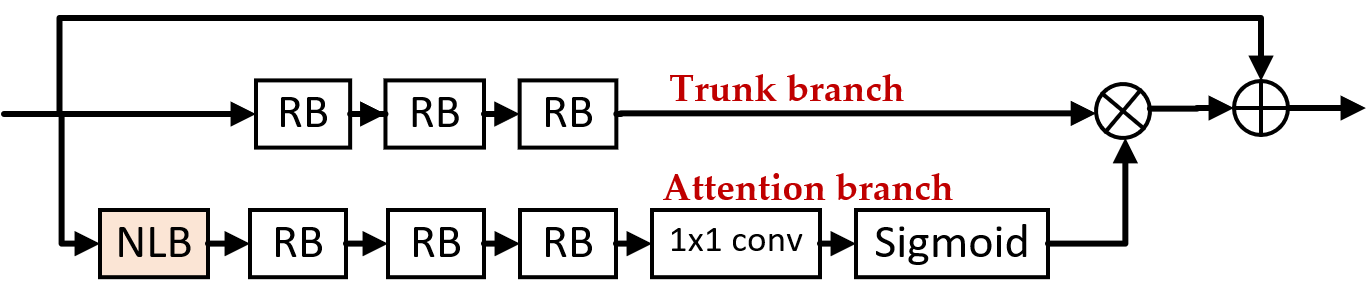}}
\subfigure[The attention module, similar to~\cite{IEEEexample:Tucodec}]{
\label{Fig.attention.2}
\includegraphics[width = 8.4 cm]{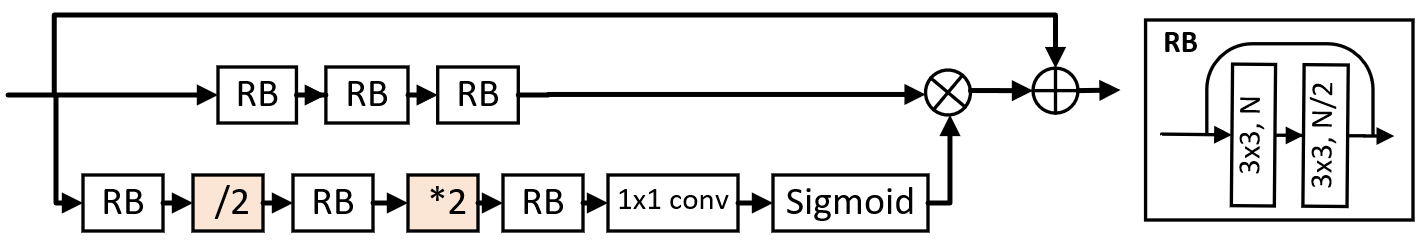}}
\subfigure[A Simplified attention module we used]{
\label{Fig.attention.3}
\includegraphics[width = 8.5cm]{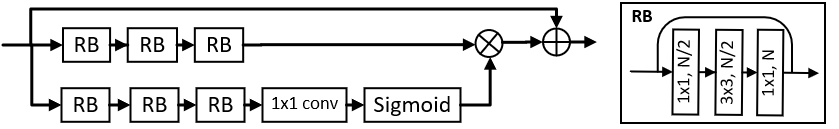}}
\caption{Different attention modules.}
\label{fig:attention}
\end{figure}

Attention module can increase the values of responses which are originally large and decrease the values of response which are originally small, thus it forces models to pay more attention to complex regions instead of simple regions to improve the coding performance, indicated by~\cite{IEEEexample:HaojieLiu, IEEEexample:Tucodec}, although the structures of attention modules are slightly different as shown in Fig.~\ref{fig:attention}. By experiments, we find non-local block (NLB), proposed by~\cite{IEEEexample:NLB} and used in~\cite{IEEEexample:HaojieLiu} is time-consuming for training and also memory-consuming when the resolution of input image is very large during inference. The work~\cite{IEEEexample:Tucodec} used a variant of attention module as Fig.~\ref{Fig.attention.2} by removing NLBs, but introduced a pair of downscale and upscale convolution in the attention branch motivated from~\cite{IEEEexample:ICLR2019}. The key point is to grasp information with larger receptive field size and large-stride convolution can increase receptive field to obtain more sophisticated attention map and capture long-range dependencies for image restoration task. Because our network for image compression already used deep residual blocks to capture large enough receptive field, so we used a more simplified version as Fig.~\ref{Fig.attention.3}. Different from~\cite{IEEEexample:HaojieLiu, IEEEexample:Tucodec}, we also modified the residual block in attention modules with 1x1x$\frac{N}{2}$-3x3x$\frac{N}{2}$-1x1x$N$ to replace 3x3x$N$-3x3x$N$ which they used, to avoid too much overhead of increasing number of parameters. Then we insert our simplified attention module into encoder-decoder network as Fig.~\ref{fig:net}.


\subsection{Discretized Gaussian Mixture Model}

Following the work~\cite{IEEEexample:CVPR2020}, we utilize discretized Gaussian mixture likelihoods to replace single Gaussian model. The motivation is to consider more flexible parameterized distributions to achieve arbitrary likelihoods, to fully utilize the contexts and information from neighboring elements and additional bits $\boldsymbol{\hat{z}}$. The Gaussian mixture model is formulated by
\begin{equation}
\label{eq.continuous}
p_{\hat{\boldsymbol{y}}|\hat{\boldsymbol{z}}}(\hat{\boldsymbol{y}}|\hat{\boldsymbol{z}}) \sim \sum_{k=1}^{K}\boldsymbol{w}^{(k)}\mathcal{N}(\boldsymbol{\mu}^{(k)}, \boldsymbol{\sigma}^{2(k)})
\end{equation}
where $\hat{\boldsymbol{y}}$ is discrete-valued after quantization. The reason why we did not use Logistic mixture likelihoods is that Gaussian achieves slightly better performance than logistic~\cite{IEEEexample:David}. Then the entropy model in end-to-end learned image compression is calculated as
\begin{equation}
\begin{aligned}
p_{\hat{\boldsymbol{y}}|\hat{\boldsymbol{z}}}(\hat{\boldsymbol{y}}|\hat{\boldsymbol{z}}) &= \prod_{i} p_{\hat{\boldsymbol{y}}|\hat{\boldsymbol{z}}}(\hat{y}_{i}|\hat{\boldsymbol{z}})\\
p_{\hat{\boldsymbol{y}}|\hat{\boldsymbol{z}}}(\hat{y}_{i}|\hat{\boldsymbol{z}}) &= (\sum_{k=1}^{K}w_{i}^{(k)}\mathcal{N}(\mu_{i}^{(k)}, \sigma_{i}^{2(k)})\ast \mathcal{U}(-\frac{1}{2}, \frac{1}{2}))(\hat{y}_{i})\\
&=\sum_{k=1}^{K}w_{i}^{(k)}(c^{(k)}(\hat{y}_{i} + \frac{1}{2}) - c^{(k)}(\hat{y}_{i} - \frac{1}{2}))
\end{aligned}
\end{equation}
where $i$ specifies the location in feature maps, and $k$ denotes the index of mixtures. Each mixture is characterized by a Gaussian distribution with $3$ parameters, i.e. weights $w_{i}^{(k)}$, means $\mu_{i}^{(k)}$ and variances $\sigma_{i}^{2(k)}$ for each element $\hat{y}_{i}$ and weights are normalized by passing through a softmax layer. $c^{(k)}$ is the cumulative distribution function for each mixture. The range of $\boldsymbol{\hat{y}}$ is automatically learned and unknown ahead of time. To achieve stable training, we clip the range of $\hat{\boldsymbol{y}}$ to [-255, 256] because empirically $\hat{\boldsymbol{y}}$ would not exceed this range. For the edge case of $-255$, replace $c(\hat{y}_{i} - \frac{1}{2})$ by zero, i.e. $c(-\infty)=0$. For the edge case of $256$, replace $c(\hat{y}_{i} + \frac{1}{2})$ by one, i.e. $c(+\infty)=1$. It provides a numerically stable implementation for training.

\section{Implementation Details and Results}

For training, we used a subset of OpenImage database~\cite{IEEEexample:OpenImage} and CLIC train dataset~\cite{IEEEexample:CLICdata}. To train our image compression models, the model was optimized using Adam~\cite{IEEEexample:adam} with a batch size of 8. $N$ is set as $128$ for low bitrate models. The learning rate was maintained at a fixed value of $1\times10^{-4}$ during the training process, and was reduced to $1\times10^{-5}$ for the last $80k$ iterations. Each model was trained to a total of $1.8\times10^{6}$ iterations for each $\lambda$ to achieve stable performance.

We optimized our models using MS-SSIM quality metrics~\cite{IEEEexample:msssim} to achieve better visual quality and distortion term is defined by $\mathcal{D}(\boldsymbol{x}, \hat{\boldsymbol{x}}) = 1 - \rm{MS\text{-}SSIM}(\boldsymbol{x}, \hat{\boldsymbol{x}})$, where the weights in mult-scale SSIM is defined as the default values [0.0448, 0.2856, 0.3001, 0.2363, 0.1333]. Finally, the loss function is defined as
\begin{equation}
\begin{aligned}
\mathcal{L} =& \mathcal{R(\hat{\boldsymbol{y}})} + \mathcal{R(\hat{\boldsymbol{z}})} + \lambda\cdot\mathcal{D(\boldsymbol{x}, \boldsymbol{\hat{x}})}\\
=& \mathop{\mathbb{E}}[-\log_{2}(p_{\hat{\boldsymbol{y}}|\hat{\boldsymbol{z}}}(\hat{\boldsymbol{y}}|\hat{\boldsymbol{z}}  ))] + \mathop{\mathbb{E}}[-\log_{2}(p_{\hat{\boldsymbol{z}}|\boldsymbol{\psi}}(\hat{\boldsymbol{z}}|\boldsymbol{\psi} ))]  \\
&+\lambda\cdot\mathcal{D(\boldsymbol{x}, \boldsymbol{\hat{x}})}
\end{aligned}
\end{equation}



\subsection{Acceleration strategy}

To make the autoregressive model faster during the decoding, we apply two acceleration strategies referring to~\cite{IEEEexample:zhoujing}. The first strategy is to use $5\times5$ window to feed in the context model when decoding. The mask convolution needs sequence decoding, while each time only $5\times5$ centered at a specific point is needed to update the value of $\hat{y}$ at this point, instead of feeding the whole size of $\hat{y}$ to the network. The second strategy is to add some flags to denote all-zero channels. For 0.15bpp, we have found many channels are quantized to all zeros. Therefore, we can skip the arithmetic coding for these all-zero channels to save time. The overhead bit is only $N$ bits, and in our case N is equal to 128, so only a total of 16 bytes per image is required.

\subsection{Bit optimization with the rate constraint}

To reach the rate constraint of 0.15bpp, we have trained four models with $\lambda$ in the set $I$ of $\{4.5, 6, 10, 14\}$ to increase the flexibility. The results with single model are shown in Table~\ref{Table.results}.

\begin{table}[h]
\begin{center}
\caption{Results on CLIC validation dataset~\cite{IEEEexample:CLICdata}.}
\label{Table.results}
\begin{tabular}{|l|l|l|}
 \hline
 \textbf{$\mathbf{\lambda}$} & \textbf{MS-SSIM}  & \textbf{Rate (bpp)}  \\
 \hline
 4.5   &0.9716	&0.1254    \\
 \hline
 6     &0.9755	  &0.1487 \\
 \hline
 10    &0.9813	&0.1999 \\
 \hline
 14    &0.9845	&0.2424 \\
 \hline
\end{tabular}
\end{center}
\end{table}

Then we formulate this problem as a multiple-choice knapsack problem, and solved it by using dynamic programming.
\begin{equation}
\text{max}_{\lambda\in I} \sum_{i}^{N}\text{MS-SSIM} \quad s.t. \sum_{i}^{N} R_{i,\lambda} \leq R_\text{Thre.}
\end{equation}

\begin{figure}

\pgfplotsset{
}
\pgfplotsset{every axis plot/.append style={line width=0.7pt}}
\tikzset{every mark/.append style={scale=0.5}}
\begin{tikzpicture}
\begin{axis}[
  width=7.2cm, height=5.5cm,
  x label style={at={(axis description cs:0.5,-0.1)},anchor=north},
  y label style={at={(axis description cs:-0.15,.5)},anchor=south},
  y tick label style = { /pgf/number format/.cd, precision=3, /tikz/.cd},
  xlabel = {Rate (bpp)},
  ylabel = {MS-SSIM},
  xmin = 0.10,
  xmax = 0.25,
  ymin = 0.964,
  ymax = 0.986,
  minor y tick num = 2,
  grid = both,
  grid style = {gray!30},
  legend entries = {\emph{Kattolab}, \emph{Kattolab-single model}, CVPR'20\cite{IEEEexample:CVPR2020}, ICLR'19~\cite{IEEEexample:Lee} },
  legend style={font=\fontsize{8}{8}\selectfont, row sep=-3pt},
  legend cell align=left,
  legend pos = {south east},
  ]
  \addplot[smooth, orange, mark=square*, mark size=3.5pt] table {data_CLIC_Final_MSSSIM.dat};
  \addplot[smooth, green!60!black, mark=square*, mark size=3.5pt] table {data_CLIC_CLIC2020_MSSSIM.dat};
  \addplot[smooth, blue!70!white] table {data_CLIC_CVPR2020_MSSSIM.dat};

  \addplot[smooth, gray] table {data_CLIC_Lee_MSSSIM.dat};

\end{axis}
\end{tikzpicture}
\caption{Performance Comparison on CLIC Validation.}
\label{fig:clic}

\end{figure}
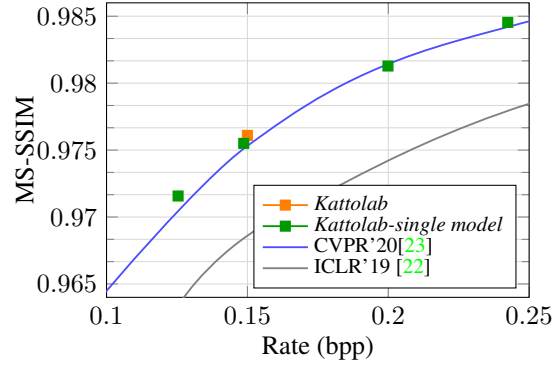

After bit allocation, MS-SSIM reaches 0.9761 at the rate of 0.15bpp. Because our submitted method is mainly based on~\cite{IEEEexample:CVPR2020}, so we also list the RD curve comparisons as Fig.~\ref{fig:clic}. Result of \emph{Kattolab} is equal to the original results of~\cite{IEEEexample:CVPR2020}.

\section{Conclusion}

In this paper, we have described our method \emph{Kattolab} for challenge on learned image compression (CLIC) 2020, which includes the network architecture, Gaussian mixture model, acceleration strategy and implementation details. Results have shown our approaches achieve 0.9761 and 0.9802 of MS-SSIM at the rate of 0.15 \emph{bpp} during the validation phase and test phase, respectively.


{\footnotesize
\begin{thebibliography}{1}

\bibitem{IEEEexample:JPEG}
G. K Wallace, \emph{``The JPEG still picture compression standard''}, IEEE Trans. on Consumer Electronics, vol. 38, no. 1, pp. 43-59, Feb. 1991.

\bibitem{IEEEexample:JPEG2000}
Majid Rabbani, Rajan Joshi, \emph{``An overview of the JPEG2000 still image compression standard''}, ELSEVIER Signal Processing: Image Communication, vol. 17, no, 1, pp. 3-48, Jan. 2002.

\bibitem{IEEEexample:HEVC}
G. J. Sullivan, J. Ohm, W. Han and T. Wiegand, \emph{``Overview of the High Efficiency Video Coding (HEVC) Standard''}, IEEE Transactions on Circuits and Systems for Video Technology, vol. 22, no. 12, pp. 1649-1668, Dec. 2012.

\bibitem{IEEEexample:VVC}
G. J. Sullivan and J. R. Ohm, \emph{``Versatile video coding Towards the next generation of video compression''}, Picture Coding Symposium, Jun. 2018.

\bibitem{IEEEexample:Theis}
Lucas Theis, Wenzhe Shi, Andrew Cunninghan and Ferenc Huszar, \emph{``Lossy Image Compression with Compressive Autoencoders''}, Intl. Conf. on Learning Representations (ICLR), pp. 1-19, April 24-26, 2017.

\bibitem{IEEEexample:Balle}
J. Ball\'e, Valero Laparra, Eero P. Simoncelli, \emph{``End-to-End Optimized Image Compression''}, Intl. Conf. on Learning Representations (ICLR), pp. 1-27, April 24-26, 2017.

\bibitem{IEEEexample:softQuan}
E. Agustsson, F. Mentzer, M. Tschannen, L. Cavigelli, R. Timofte, L. Benini, L. V. Gool, \emph{``Soft-to-Hard Vector Quantization for End-to-End Learning Compressible Representations''}, Neural Information Processing Systems (NIPS) 2017, arXiv:1704.00648v2.

\bibitem{IEEEexample:Toderici01}
G. Toderici, S. M.O'Malley, S. J. Hwang, et al., \emph{``Variable rate image compression with recurrent neural networks''}, arXiv: 1511.06085, 2015.

\bibitem{IEEEexample:Toderici}
G, Toderici, D. Vincent, N. Johnson, et al., \emph{``Full Resolution Image Compression with Recurrent Neural Networks''}, IEEE Conf. on Computer Vision and Pattern Recognition (CVPR), pp. 1-9, July 21-26, 2017.

\bibitem{IEEEexample:Nick}
Nick Johnson, Damien Vincent, David Minnen, et al., \emph{``Improved Lossy Image Compression with Priming and Spatially Adaptive Bit Rates for Recurrent Networks''}, arXiv:1703.10114, pp. 1-9, March 2017.


\bibitem{IEEEexample:waveone}
Ripple Oren, L. Bourdev, \emph{``Real Time Adaptive Image Compression''}, Proc. of Machine Learning Research, Vol. 70, pp. 2922-2930, 2017.

\bibitem{IEEEexample:MITgan}
S. Santurkar, D. Budden, N. Shavit, \emph{``Generative Compression''}, Picture Coding Symposium, June 24-27, 2018.

\bibitem{IEEEexample:Extreme}
E. Agustsson, M. Tschannen, F. Mentzer, R. Timofte, and L. V. Gool, \emph{``Generative Adversarial Networks for Extreme Learned Image Compression''}, arXiv:1804.02958.





\bibitem{IEEEexample:HKPU}
M. Li, W. Zuo, S. Gu, D. Zhao, D. Zhang, \emph{``Learning Convolutional Networks for Content-weighted Image Compression''}, IEEE Conf. on Computer Vision and Pattern Recognition (CVPR), June 17-22, 2018.


\bibitem{IEEEexample:conditional}
F. Mentzer, E. Agustsson, M. Tschannen, R. Timofte, L. V. Gool, \emph{``Conditional Probability Models for Deep Image Compression''}, IEEE Conf. on Computer Vision and Pattern Recognition (CVPR), June 17-22, 2018.


\bibitem{IEEEexample:PCS2018}
Z. Cheng, H. Sun, M. Takeuchi, J. Katto, \emph{``Deep Convolutional AutoEncoder-based Lossy Image Compression''}, Picture Coding Symposium, pp. 1-5, June 24-27, 2018.

\bibitem{IEEEexample:CLIC2018cheng}
Z. Cheng, H. Sun, M. Takeuchi, J. Katto, \emph{``Performance Comparison of Convolutional AutoEncoders, Generative Adversarial Networks and Super-Resolution for Image Compression''}, CVPR Workshop and Challenge on Learned Image Compression (CLIC), pp. 1-4, June 17-22, 2018.

\bibitem{IEEEexample:TMM2019cheng}
Z. Cheng, H. Sun, M. Takeuchi, J. Katto, \emph{``Energy Compaction-Based Image Compression Using Convolutional AutoEncoder''}, IEEE Transactions on Multimedia, vol.22, no. 4, April 2020.


\bibitem{IEEEexample:CVPR2019}
Z. Cheng, H. Sun, M. Takeuchi, J. Katto, \emph{``Learning Image and Video Compression through Spatial-Temporal Energy Compaction''}, IEEE Conf. on Computer Vision and Pattern Recognition (CVPR), 2019. arXiv.1906.09683


\bibitem{IEEEexample:Balle2}
J. Ball\'e, D. Minnen, S. Singh, S. J. Hwang, N. Johnston, \emph{``Variational Image Compression with a Hyperprior''}, Intl. Conf. on Learning Representations (ICLR), 2018.




\bibitem{IEEEexample:David}
D. Minnen, J. Ball\'{e}, G. Toderici, \emph{``Joint Autoregressive and Hierarchical Priors for Learned Image Compression''}, NeurIPS 2018, arXiv.1809.02736.

\bibitem{IEEEexample:Lee}
J. Lee, S. Cho, S-K Beack, \emph{``Context-Adaptive Entropy Model for End-to-end optimized Image Compression''}, Intl. Conf. on Learning Representations (ICLR) 2019.


\bibitem{IEEEexample:CVPR2020}
Z. Cheng, H. Sun, M. Takeuchi, J. Katto, \emph{``Learned Image Compression with Discretized Gaussian Mixture Likelihoods and Attention Modules''}, IEEE Conf. on Computer Vision and Pattern Recognition (CVPR), 2020. arXiv.2001.01568.


\bibitem{IEEEexample:CLIC2019}
Z. Cheng, H. Sun, M. Takeuchi, J. Katto, \emph{``Deep Residual Learning for Image Compression''}, CVPR Workshop, pp. 1-4, June 16-20, 2019.

\bibitem{IEEEexample:Balle3}
J. Ball\'e, \emph{``Efficient Nonlinear Transforms for Lossy Image Compression''}, Picture Coding Symposium, 2018.

\bibitem{IEEEexample:HaojieLiu}
H. Liu, T. Chen, P. Guo. Q. Shen, X. Cao, Y. Wang, Z. Ma, \emph{``Non-local Attention Optimized Deep Image Compression''}, arXiv.1904.09757.

\bibitem{IEEEexample:Tucodec}
L. Zhou, Z. Sun, X. Wu, J. Wu, \emph{``End-to-end Optimized Image Compression with Attention Mechanism''}, CVPRW 2019.


\bibitem{IEEEexample:NLB}
X. Wang, R. Girshick, A. Gupta, K. He, \emph{``Non-local Neural Networks''}, IEEE Conf. on Computer Vision and Pattern Recognition (CVPR), pp. 7794-7803, 2018.

\bibitem{IEEEexample:ICLR2019}
Y. Zhang, K. Li, K. B. Zhong, Y. Fu, \emph{``Residual non-local attention networks for image restoration''}, Intl. Conf. on Learning Representations (ICLR) 2019.


\bibitem{IEEEexample:zhoujing}
J. Zhou, S. Wen, A. Nakagawa, K. Kazui, Z. Tan, \emph{``Multi-scale and Context-adaptive Entropy Model for Image Compression''}, CVPR workshop 2019.



\bibitem{IEEEexample:OpenImage}
OpenImages, dataset available from
\url{https://storage.googleapis.com/openimages/web/download.html}.

\bibitem{IEEEexample:CLICdata}
Workshop and Challenge on Learned Image Compression (CLIC), \url{http://www.compression.cc/}



\bibitem{IEEEexample:adam}
D. P. Kingma and J. Ba, \emph{``Adam: A method for stochastic optimization''}, arXiv:1412.6980, pp.1-15, Dec. 2014.

\bibitem{IEEEexample:msssim}
Z. Wang, E. P. Simoncelli and A. C. Bovik, \emph{``Multiscale structural similarity for image quality assessment''}, The 36-th Asilomar Conference on Signals, Systems and Computers, Vol.2, pp. 1398-1402, Nov. 2013.



\end{thebibliography}
}

\end{document}